\documentclass[english,letterpaper,showpacs,floatfix]{revtex4}

\usepackage{epsfig}
\usepackage{amsfonts}
\usepackage{amsmath}

\begin{document}

\title{Classical Non-Local conserved charges in String Theory}
\author{Elcio Abdalla}
\email{eabdalla@fma.if.usp.br}
\affiliation{Instituto De Fisica, Universidade De Sao Paulo, C.P.66.318,
CEP 05315-970, Sao Paulo, Brazil}
\author{Antonio Lima-Santos}
\email{dals@df.ufscar.br}
\affiliation{Universidade Federal de S\~ao Carlos, Departamento de
F\'{\i}sica \\ 
Caixa Postal 676, 13569-905, S\~ao Carlos-SP, Brazil}

\begin{abstract}
We construct a conserved non local charge in $AdS_5\times S_5$ string
theory. 
\end{abstract}

\maketitle


String theory is the strongest candidate to describe all interactions\cite%
{gsw}. However, there is no crucial experiment giving full support of the
theory which is still in need of a stronger basis for a full theory of
nature. As happens in several field theories, non perturbative results may
give important clues to the behaviour of the theory in particularly
difficult situations. Several results have already been obtained from the
idea of duality \cite{duality}. Recently some new features connected with
the high dimensionality of strings have led to further insight into the
structure of the socalled brane cosmology \cite%
{wittenvariousdim,horavawitten,rs}.

More recently, some authors are pursuing higher conservation laws \cite%
{roiban,nappi,breno}, which proved of great help in theories of lower
dimensionality \cite{grossneveu,luescher,abdalla,aar}. In case we can use
higher conservation laws in a way similar to the one used in two dimensional
space time, it is possible that further constraints in the dynamical
behaviour of strings can be imposed and one can gather information based on
more general grounds to be compared with observations. As an example, strong
nonperturbative insight about gravity can be obtained from the holographic
principle \cite{holo}.


We begin our problem with a set of currents defined in $AdS_5\times S_5$ space
described by the coset $PSU(2,2\vert 4)/SO(4,1)\times SO(5)$
\cite{metsaev}. The underlying string theoryhas been described accordingly
\cite{nathan}. The algebra $psu(2,2\vert 4)$ has, under the discrete group
$Z_4$, a discrete decomposition $\mathcal{H}=
\oplus_{i=0}^{3}\mathcal{H}_i$, described by
\begin{eqnarray}
t_{ \alpha}\in \mathcal{H}_1\quad ,\qquad t_{\underline a}\in \mathcal{H}_2 
\quad , \notag \\
t_{\hat \alpha}\in \mathcal{H}_3\quad ,\qquad t_{\underline{\lbrack a
b\rbrack}}\in \mathcal{H}_0  \label{deco}
\end{eqnarray}
where $\underline a$ are indices parametrizing $AdS_5\times S_5$, $\alpha$ and 
$\hat\alpha$ are the superspace connections. The non vanishing structure
constants are well known \cite{breno}.

In terms of the supercoset valued filed $g(x,\theta ,\hat{\theta})$ we can
define algebra valued currents $\mathbf{J}=g^{-1}dg$, which in turn are can
be decomposed according to (\ref{deco}). The resulting currents have been
shown to obey the relations \cite{breno}%
\begin{eqnarray}
\partial _{\mu }J_{1}^{\mu }+\left[ J_{0\mu },J_{1}^{\mu }\right] 
&=&\varepsilon ^{\mu \nu }\left[ J_{2\mu },J_{3\nu }\right] \quad ,  \notag \\
\partial _{\mu }J_{2}^{\mu }+\left[ J_{0\mu },J_{2}^{\mu }\right]  &=&\frac{1%
}{2}\varepsilon ^{\mu \nu }\left( \left[ J_{3\mu },J_{3\nu }\right] -\left[
J_{1\mu },J_{1\nu }\right] \right)   \quad ,  \notag \\
\partial _{\mu }J_{3}^{\mu }+\left[ J_{0\mu },J_{3}^{\mu }\right] 
&=&-\varepsilon ^{\mu \nu }\left[ J_{2\mu },J_{1\nu }\right]  \quad ,   
\label{div}
\end{eqnarray}%
and 
\begin{eqnarray}
\varepsilon ^{\mu \nu }\left( \partial _{\mu }J_{1\nu }+\left[ J_{0\mu
},J_{1\nu }\right] \right)  &=&-\varepsilon ^{\mu \nu }\left[ J_{2\mu
},J_{3\nu }\right]  \quad ,   \notag \\
\varepsilon ^{\mu \nu }\left( \partial _{\mu }J_{2\nu }+\left[ J_{0\mu
},J_{2\nu }\right] \right)  &=&-\frac{1}{2}\varepsilon ^{\mu \nu }\left( %
\left[ J_{3\mu },J_{3\nu }\right] +\left[ J_{1\mu },J_{1\nu }\right]
\right)  \quad , \notag \\
\varepsilon ^{\mu \nu }\left( \partial _{\mu }J_{3\nu }+\left[ J_{0\mu
},J_{3\nu }\right] \right)  &=&-\varepsilon ^{\mu \nu }\left[ J_{2\mu
},J_{1\nu }\right]   \quad ,  \label{rot}
\end{eqnarray}%
in the underlying Minkowski space. Such relations keep some similarity
with zero curvature conditions, well known in integrable models
\cite{grossneveu} and nonlinear sigma models \cite{luescher,abdalla}, 
implying, upon quantization, severe
constraints upon the S-matrix elements \cite{aar}.

Here, we try a constructive approach. First, we neglect terms related to
gauge-field valued elements of the algebra, which simplifies the discussion.
In fact, the relevant currents transform nontrivially under gauge
transformations, which is mirrored in the fact that the derivatives in the
conservation laws are covariant derivatives of the form $\partial +[\mathbf{J%
}_{0},]$. We also notice that some commutators are gauge valued, such as $[%
\mathbf{J}_{2},\mathbf{J}_{2}]$, or $[\mathbf{J}_{1},\mathbf{J}_{3}]$. In
expression (\ref{div}), there are several conservation laws inbuilt
and we are going to construct one of them. We claim that a non local
conserved charge should be described in terms of a gauge dressing of a
combination of the following building blocks:
\begin{eqnarray}
Q^{(1)} &=&2\int J_{3}^{0}(t,x_{1})dx_{1} \quad ,  \notag \\
Q^{(2)} &=&-\int J_{1}^{0}(t,x_{1})\epsilon
(x_{1}-x_{2})J_{2}^{0}(t,x_{2})dx_{1}dx_{2}-\int J_{2}^{0}(t,x_{1})\epsilon
(x_{1}-x_{2})J_{1}^{0}(t,x_{2})dx_{1}dx_{2} \quad ,  \notag \\
Q^{(3)} &=&\frac{1}{2}\int J_{1}^{0}(t,x_{1})\epsilon
(x_{1}-x_{2})J_{1}^{0}(t,x_{2})\epsilon
(x_{2}-x_{3})J_{1}^{0}(t,x_{3})dx_{1}dx_{2}dx_{3}   \notag \\
&-&\frac{1}{2}\int J_{3}^{0}(t,x_{1})\epsilon
(x_{1}-x_{2})J_{3}^{0}(t,x_{2})\epsilon
(x_{2}-x_{3})J_{1}^{0}(t,x_{3})dx_{1}dx_{2}dx_{3} \notag \\
&-&\frac{1}{2}\int J_{3}^{0}(t,x_{1})\epsilon
(x_{1}-x_{2})J_{1}^{0}(t,x_{2})\epsilon
(x_{2}-x_{3})J_{3}^{0}(t,x_{3})dx_{1}dx_{2}dx_{3}  \notag \\
&-&\frac{1}{2}\int J_{1}^{0}(t,x_{1})\epsilon
(x_{1}-x_{2})J_{3}^{0}(t,x_{2})\epsilon
(x_{2}-x_{3})J_{3}^{0}(t,x_{3})dx_{1}dx_{2}dx_{3}  \notag \\
&-&\frac{1}{2}\int J_{2}^{0}(t,x_{1})\epsilon
(x_{1}-x_{2})J_{2}^{0}(t,x_{2})\epsilon
(x_{2}-x_{3})J_{3}^{0}(t,x_{3})dx_{1}dx_{2}dx_{3}  \notag \\
&-&\frac{1}{2}\int J_{2}^{0}(t,x_{1})\epsilon
(x_{1}-x_{2})J_{3}^{0}(t,x_{2})\epsilon
(x_{2}-x_{3})J_{2}^{0}(t,x_{3})dx_{1}dx_{2}dx_{3}  \notag \\
&-&\frac{1}{2}\int J_{3}^{0}(t,x_{1})\epsilon
(x_{1}-x_{2})J_{2}^{0}(t,x_{2})\epsilon
(x_{2}-x_{3})J_{2}^{0}(t,x_{3})dx_{1}dx_{2}dx_{3} \quad ,  \notag \\
Q^{(4)} &=&\frac{1}{4}\sum \int J_{1}^{0}(x_{1})\epsilon
(x_{1}-x_{2})J_{1}^{0}(x_{2})\epsilon (x_{2}-x_{3})J_{2}^{0}(x_{3})\epsilon
(x_{3}-x_{4})J_{3}^{0}(x_{4})dx_{1}dx_{2}dx_{3}dx_{4}  \notag \\
&+&\frac{1}{4}\sum \int J_{1}^{0}(x_{1})\epsilon
(x_{1}-x_{2})J_{2}^{0}(x_{2})\epsilon (x_{2}-x_{3})J_{2}^{0}(x_{3})\epsilon
(x_{3}-x_{4})J_{2}^{0}(x_{4})dx_{1}dx_{2}dx_{3}dx_{4}  \notag \\
&-&\frac{1}{4}\sum \int J_{2}^{0}(x_{1})\epsilon
(x_{1}-x_{2})J_{3}^{0}(x_{2})\epsilon (x_{2}-x_{3})J_{3}^{0}(x_{3})\epsilon
(x_{3}-x_{4})J_{3}^{0}(x_{4})dx_{1}dx_{2}dx_{3}dx_{4}\quad ,   \notag \\
&&...
\end{eqnarray}%
where the sum is over all orders of the indices of currents. The remaining
terms are to be constructed taking into account the generic additive term%
\begin{equation}
Q_{i}^{(n)}=\pm \frac{1}{2^{n-2}}\sum \int \left(
\prod_{k=1}^{n-1}J_{\alpha _{k}}^{0}(t,x_{k})\epsilon
(x_{k}-x_{k+1})dx_{k}\right) J_{\alpha _{n}}^{0}(t,x_{n})dx_{n},
\end{equation}%
with the indices $\alpha _{k}$ satisfying the constraint equation%
\begin{equation}
\sum_{k=1}^{n}\alpha _{k}=3\quad \mod 4.
\end{equation}
The sign has to be properly chosen in order to achieve conservation of the
sum. The time derivative of the current can be exchanged by the space
derivative, a commutator with a gauge-valued current and a nontrivial 
commutator. The
space derivative is integrated by parts giving rise to a Dirac delta used to
perform one integration, and leaving a lower order term in the integration
variables, but with a nontrivial commutator. The commutator either cancels a
similar one arising from a lower order derivative, or is gauge valued.
Therefore, we claim that 
\begin{equation}
\partial _{0}Q^{(n)}=\delta _{der}Q^{(n)}+\delta _{com}Q^{(n)}+{%
\hbox{gauge
terms}}
\end{equation}%
where the first term arises from the integration by parts of the space
derivative of the current giving rise to a Dirac-delta term of the form of a
commutator among two currents connected by the $\epsilon (x_{k}-x_{k+1})$
function, while the second term arises from the remaining term of the
equation of motion, with exception of the gauge ($\mathbf{J}_{0}$)
commutators. In the first term we leave aside the commutators which take
values in the gauge sector, namely, $[\mathbf{J}_{2},\mathbf{J}_{2}]$ and $[%
\mathbf{J}_{1},\mathbf{J}_{3}]$, which are, together with the explicit gauge
terms containing $\mathbf{J}_{0}$ left to the last term. Under these
conditions, the definition of the charges leads us to the result 
\begin{equation}
\delta _{der}Q^{(n+1)}=-\delta _{com}Q^{(n})\quad .
\end{equation}%
Therefore, we

\textit{Claim}: the nonlocal charge 
\begin{equation}
Q_{3}=\sum_{n=1}^{\infty }Q^{(n)}
\end{equation}%
is classically conserved up to a gauge dressing, defined substituting some
currents by $\mathcal{H}_{0}$ valued elements.

We also claim that analogous charges, whose first term are either obtained
from $J_{2}^{0}$ or from $J_{1}^{0}$ are also conserved.

\textbf{Acknowledgement}: This work has been supported by CNPq and FAPESP,
Brazil. We would like like to thank several discussions with Dr. Breno
Vallilo. 



\begin{thebibliography}{99}
\bibitem{gsw} M. Green, J. Schwarz and E. Witten \textit{Superstring Theory}%
, Cambridge Univ. Pr. (1987), Cambridge Monographs On Mathematical Physics. 

\bibitem{duality} J. Polchinski \textit{Superstring Theory}, Cambridge
University Press 1998. 

\bibitem{wittenvariousdim} E. Witten \textit{Nucl. Phys.} \textbf{B443}
(1995) 85, hep-th/9503124. 

\bibitem{horavawitten} P. Horava and E. Witten \textit{Nucl. Phys.} \textbf{%
\ B460} (1996) 506, \textbf{B475} (1996) 94. 

\bibitem{rs} L. Randall and R. Sundrum \textit{Phys. Rev. Lett.} \textbf{83}
(1999) 3370, 4690. 

\bibitem{roiban} Iosif Bena, Joseph Polchinski, Radu Roiban \textit{\ Phys.
Rev.} \textbf{D69} (2004) 046002, hep-th/0305116; Bin Chen, Xiao-Jun Wang
and Yong-Shi Wu \textit{Phys. Lett.} \textbf{B591} (2004) 170
hep-th/0403004, \textit{JHEP} \textbf{0402} (2004) 029, hep-th/0401016. 

\bibitem{nappi} L. Dolan, C. Nappi and E. Witten \textit{JHEP} \textbf{0310}
(2003) 017, hep-th/0308089. 

\bibitem{breno} Brenno Carlini Vallilo, \textit{JHEP} \textbf{0403} (2004)
037, hep-th/0307018. 

\bibitem{grossneveu} D. Gross, Neveu \textit{Phys. Rev.} \textbf{D10}
(1974)3235; E. Abdalla and A. Lima-Santos \textit{Rev. Bras. Fis.} \textbf{12%
} (1982) 293. 

\bibitem{luescher} M. L\"uscher \textit{Nucl. Phys.} \textbf{B135} (1978) 1;
M. Luscher and K. Pohlmeyer \textit{Nucl. Phys.} \textbf{B137} (1978) 46. 

\bibitem{abdalla} E. Abdalla, M. C. B. Abdalla and M. Gomes \textit{\ Phys.
Rev.}\textbf{D23} (1981) 1800, \textbf{D25} (1982) 452, \textbf{D27} (1983)
825; E. Abdalla, M. Forger and M. Gomes \textit{\ Nucl. Phys.} \textbf{B210}
[FS6] (1982) 181; E. Abdalla, M. Forger and A. Lima-Santos \textit{Nucl.
Phys.} \textbf{B256} (1985) 145; E. Abdalla, M. C. B. Abdalla and M. Forger 
\textit{\ Nucl. Phys.} \textbf{B297} (1988) 374; E. Abdalla and A.
Lima-Santos, \textit{Mod. Phys. Lett.} \textbf{A3} (1988) 310, \textit{Phys.
Rev.} \textbf{D29} (1984) 1851; E. Abdalla, M.C.B. Abdalla and A.
Lima-Santos, \textit{Phys. Lett.} \textbf{B140} (1984) 71-75, Erratum 
\textit{B146} (1984) 457. 

\bibitem{aar} E. Abdalla, M. C. B. Abdalla and K. D. Rothe, \textit{Non
perturbative methods in two dimensional quantum field theory}, World
Scientific Publishing Company, 2001. 

\bibitem{holo} Leonard Susskind \textit{J. Math. Phys.} \textbf{36} (1995)
6377 hep-th/9409089; Gerard 't Hooft \textit{Salamfest} 1993, 0284-296
(QCD161:C512:1993) gr-qc/9310026. 
\bibitem{metsaev} R. R. Metsaev and A. A. Tseytlin \textit{Nucl. Phys.} 
\textbf{\ B596} (2001) 185, hep-th/0009168. 
\bibitem{nathan} N. Berkovits {\it JHEP} {\bf 04} (2000) 018; N. Berkovits
and O. Chandia {\it Nucl. Phys.} {\bf B596} (2001) 185; B. Vallilo  {\it
JHEP} {\bf 0212} (2002) 042.
\end{thebibliography}
\end{document}